\documentclass[letterpaper, 11pt]{article}

\usepackage{graphicx}
\usepackage[dvips]{color}
\usepackage{natbib}
\usepackage{MnSymbol}

\begin{document}

\begin{center}

{\LARGE Axisymmetric viscous gravity currents\\ flowing over a porous medium}

\vspace{0.25in}

{\Large Melissa J. Spannuth$^1$, Jerome A. Neufeld$^2$,\\ J. S. Wettlaufer$^{1,3,4}$, and M. Grae Worster$^2$}

\vspace{0.25in}

{\em \footnotesize 1. Department of Geology and Geophysics, Yale University, New Haven, CT 06520, USA\\ 2. Institute of Theoretical Geophysics, Department of Applied Mathematics and Theoretical Physics,\\ University of Cambridge, Wilberforce Road, CB3 0WA, UK\\ 3. Department of Physics, Yale University, New Haven, CT 06520, USA\\ 4. Nordic Institute for Theoretical Physics, Roslagstullsbacken 23, University Center, 106 91 Stockholm, Sweden}
\end{center}

\vspace{0.25in}

\begin{abstract}
We study the axisymmetric propagation of a viscous gravity current over a deep porous medium into which it also drains. A model for the propagation and drainage of the current is developed and solved numerically in the case of constant input from a point source. In this case, a steady state is possible in which drainage balances the input, and we present analytical expressions for the resulting steady profile and radial extent. We demonstrate good agreement between our experiments, which use a bed of vertically aligned tubes as the porous medium, and the theoretically predicted evolution and steady state. However, analogous experiments using glass beads as the porous medium exhibit a variety of unexpected behaviours, including overshoot of the steady-state radius and subsequent retreat, thus highlighting the importance of the porous medium geometry and permeability structure in these systems.
\end{abstract}

\section{Introduction}

Gravity currents are primarily horizontal fluid flows driven by a density difference between the intruding and ambient fluids. These flows are common in natural systems and industrial processes and describe, for example, the spread of cold air into a room, the dispersal of pollutants from an industrial spill, and the flow of snow and debris avalanches \citep{huppert-2006}. Many previous studies have examined in detail the propagation of currents along impermeable boundaries; here we consider flow over porous substrates through which these currents can also drain.

Two-dimensional gravity currents propagating over porous media have been addressed both theoretically and experimentally by several authors. For currents flowing over thin porous substrates, only the weight of the overlying fluid drives drainage \citep*{thomas-1998, ungarish-2000, marino-2002, pritchard-2001}. In contrast, for gravity currents propagating over deep porous media, \cite*{acton-2001} showed that both the hydrostatic pressure of the fluid in the current and the weight of the fluid within the porous medium drive drainage. They used this description of drainage in a model of experiments in which low Reynolds number gravity currents spread over a deep porous layer in two dimensions. \cite*{thomas-2004} used this drainage law to describe their experiments on the propagation of high Reynolds number currents over deep porous media. \cite{pritchard-2002} have also applied the same drainage law to their examination of gravity currents propagating within a porous medium overlying a deep layer of lower permeability. Similar studies have also been conducted that consider two-phase flow within the porous medium  \cite*[e.g.][]{hussein-2002}, but these effects are beyond the scope of the present study.

Axisymmetric gravity currents propagating over porous media have been studied primarily as microscale flows in which capillary forces drive the drainage and therefore the wetting properties of the medium are important \cite[e.g.][]{davis-1999, davis-2000, kumar-2006}. At the macroscale, \cite{pritchard-2001} considered gravity-driven drainage of an axisymmetric current flowing through a porous medium overlying a thin layer of lower permeability. In both geometries, previous experiments only involved currents of fixed volume, whereas our experiments explore the fixed flux case.

Here we examine the axisymmetric propagation of a macroscopic viscous gravity current over a deep porous medium. Our model uses lubrication theory for flow within the current, the drainage law of \cite{acton-2001}, and Darcy flow within the porous medium. While the full spatial and temporal evolution of the current can only be obtained numerically, an analytical expression for the steady-state extent and profile of a current fed by a constant input of fluid is found. Additionally, we develop scaling laws describing the propagation of the current. Our experimental setup in which a gravity current fed by a constant flux of golden syrup spreads across a bed of vertically aligned straws conforms closely to the assumptions of our model, so the scaling laws provide a good collapse of all data onto a curve in agreement with the numerical solution. In contrast to these well-behaved currents, we describe the non-ideal behaviour observed in experiments using glycerin and glass beads similar to the system used by \cite{acton-2001}. We propose that the axisymmetric geometry makes the currents particularly sensitive to any non--uniformities of the porous medium which leads to the disagreement between these experiments and the theory.

\section{Theoretical model}

\begin{figure}
\centering
\includegraphics[width = 4.5in]{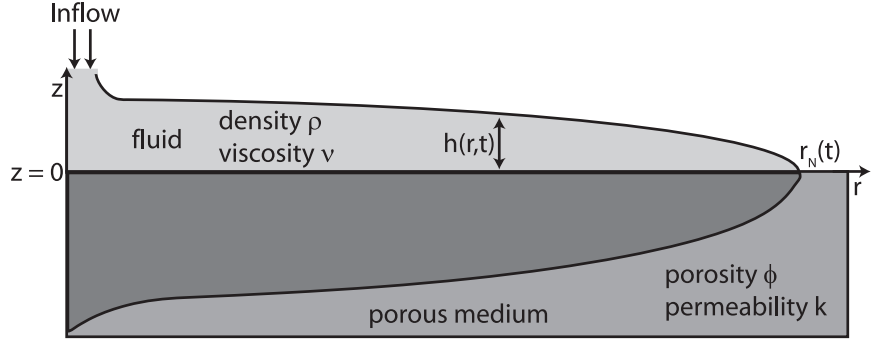}
\caption{An illustration of the theoretical and experimental geometry considered for the axisymmetric spreading of a viscous gravity current over a porous medium. \label{ASgeom}}
\end{figure}

We consider the axisymmetric spreading of a fluid of kinematic viscosity $\nu$ and density $\rho$ into an ambient fluid of density $\rho_a \ll \rho$ and viscosity $\nu_a \ll \nu$. As shown schematically in figure~\ref{ASgeom}, fluid is supplied at the origin and spreads radially over a porous medium with porosity $\phi$ and permeability $k$ into which the fluid drains. We consider the general case in which the volume of fluid increases as $q t^{\alpha}$, where $t$ is time and $q$ and $\alpha$ are constants. After a brief initial stage, the radial extent of the current $r_N(t)$ is much greater than its height $h(r,t)$, and in this limit we apply the approximations of lubrication theory: velocity within the current is assumed to be predominantly horizontal and pressure within the current is assumed to be hydrostatic. Under this approximation, viscous flow within the current is driven by radial gradients of its thickness. We apply conditions of no slip at $z = 0$ and no tangential stress at $z = h$ to determine the horizontal fluid flux $q_h = -\left( g / 3 \nu \right) \, r h^3 \, \partial h / \partial r$. Conservation of fluid mass through an infinitesimal control volume of the gravity current gives the equation
\begin{equation} 
\frac{\partial h}{\partial t} - \frac{g}{3\nu}\frac{1}{r}\frac{\partial}{\partial r}\left(rh^3\frac{\partial h}{\partial r}\right) = w(r,0,t) \label{heightevo}
\end{equation}
 governing the current's structure and evolution, where $w(r,0,t)$ is the drainage velocity from the base of the current into the underlying porous medium. Following \cite{acton-2001}, we assume that drainage into the porous medium is driven both by the weight of the draining fluid and the hydrostatic pressure of the fluid within the current, giving
\begin{equation}
w(r,0,t) = -\frac{g k}{\nu}\left(1+\frac{h}{l}\right) = -\phi \frac{\partial l}{\partial t}, \label{depthevo}
\end{equation}
where $l(r,t)$ is the depth of the fluid within the porous medium.

In this analysis, we have made a few assumptions that merit further examination. First, our assumption of no slip at the porous medium surface is valid when the current height is much greater than the pore size because the presence of a slip velocity is equivalent to extending the fluid region a distance of less than one pore size into the medium \citep*{beavers-1967,lebars-2006}. Near the nose of the current or for currents flowing over very rough substrates apparent slip may be important. Secondly, we have assumed that surface tension is negligible in the drainage law. As discussed in \cite{acton-2001}, this is accurate as long as the pressure due to surface tension is much less than the hydrostatic pressure; equivalently the capillary rise height $h_c \approx \gamma / \rho g a$, where  $\gamma$ is the surface tension and $a$ is the pore radius, must be much larger than $h$. For our experimental setup $h_c \approx 2\ \mbox{mm}$, much less than typical current heights. Finally, equation (\ref{depthevo}) assumes that flow within the porous medium is single-phase and that the porous matrix is stationary; therefore we can ignore flow of the displaced fluid and assume that $k$ is constant in time. This is an accurate assumption when the wetting properties of the displaced and displacing fluids with respect to the porous matrix are similar and when the displaced fluid is inviscid. Equation (\ref{depthevo}) also assumes that fluid flow is predominantly vertical, implicitly neglecting the potential for the Rayleigh--Taylor instability.

The governing equations (\ref{heightevo}) and (\ref{depthevo}) are subject to one boundary condition specifying the flux near the origin and another requiring zero flux through the nose of the current $r_N(t)$. Respectively, they are
\refstepcounter{equation}
$$
\lim_{r \to 0} \left[ 2\pi r \frac{g h^3}{3 \nu} \frac{\partial h}{\partial r} \right] = -\alpha q t^{\alpha-1} \label{originflux} 
\quad \mbox{and} \quad \left[ 2\pi r \frac{g h^3}{3 \nu} \frac{\partial h}{\partial r} \right]_{r_N} = 0. \label{nosedef} 
\eqno{(\theequation{\mathit{a},\mathit{b}})}
$$
We note that boundary condition~(\ref{originflux}\textit{a}) along with the evolution equation~(\ref{heightevo}) is equivalent to a statement of global mass conservation, namely
\begin{equation} 
qt^\alpha = 2\pi\int_{0}^{r_N} r h \, dr - 2\pi\int_0^t\int_{0}^{r_N} r w(r,0,t) \, dr \, dt. \label{gcons}
\end{equation}

We non-dimensionalize equations~(\ref{heightevo})--(\ref{nosedef}) by introducing horizontal, vertical and temporal scales $S_H$, $S_V$ and $S_T$ given by
\refstepcounter{equation}
$$
S_H = \left( q/\Gamma \right)^{1/2} \left( \frac{\Gamma^4 g}{3 q \nu} \right)^{\left(\alpha-1\right)/2\left(\alpha-5\right)}, \quad S_V = \Gamma \left( \frac{\Gamma^4 g}{3 q \nu} \right)^{1/\left(\alpha-5\right)} \quad\mbox{and}\quad S_T = \left( \frac{\Gamma^4 g}{3 q \nu} \right)^{1/\left(\alpha-5\right)},
\eqno{(\theequation{\mathit{a},\mathit{b},\mathit{c}})}
$$
where $\Gamma = gk/\nu$ is the characteristic drainage velocity in the porous medium. By introducing dimensionless variables
\refstepcounter{equation}
$$
H = h/S_V, \quad L = l/S_V, \quad R = r/S_H \quad \mbox{and} \quad T = t/S_T,
\eqno{(\theequation{\mathit{a},\mathit{b},\mathit{c},\mathit{d}})}
$$
the equations governing the dimensionless height $H(R,T)$ and depth $L(R,T)$ of the intruding fluid become
\begin{equation}
\frac{\partial H}{\partial T} - \frac{1}{R}\frac{\partial}{\partial R}\left(RH^3\frac{\partial H}{\partial R}\right) = -\left(1+\frac{H}{L}\right) \label{NDH}
\end{equation}
and
\begin{equation}
\phi\frac{\partial L}{\partial T} = \left(1+\frac{H}{L}\right). \label{NDL}
\end{equation}
These are subject to the scaled boundary conditions
\refstepcounter{equation}
$$
\lim_{R \to 0} \left[ 2\pi RH^3\frac{\partial H}{\partial R} \right] = -\alpha T^{\alpha-1} \label{NDoriginflux} 
\quad \mbox{and} \quad  \left[ 2\pi RH^3\frac{\partial H}{\partial R} \right]_{R_N} = 0. \label{NDnosedef}
\eqno{(\theequation{\mathit{a},\mathit{b}})}
$$

\begin{figure}
\centering
\includegraphics[width = 4.5in]{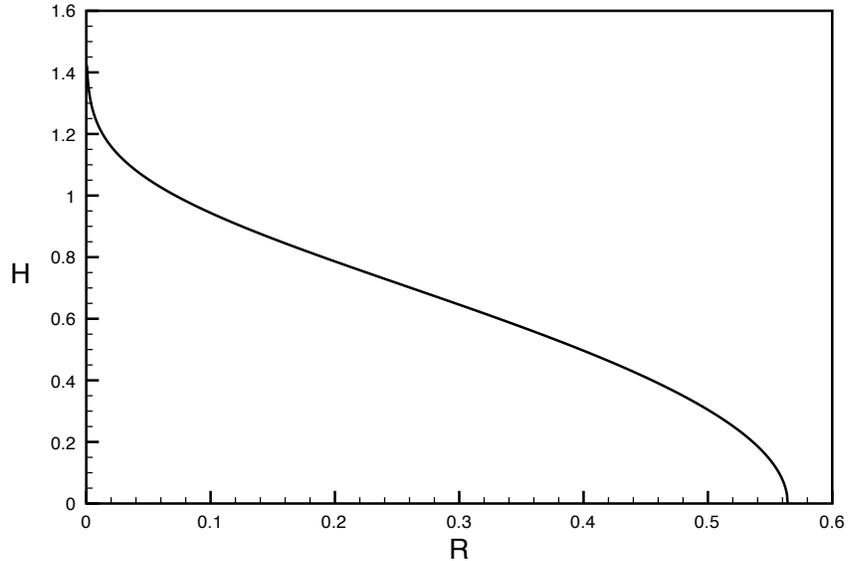}
\caption{The steady-state profile of dimensionless height $H$ versus dimensionless radius $R$. \label{ss}}
\end{figure}

For a fixed flux at the origin ($\alpha = 1$) the scalings simplify to
\refstepcounter{equation}
$$
S_H = \left(\frac{q\nu}{gk}\right)^{1/2}, \quad S_V = \left(\frac{3q\nu}{g}\right)^{1/4} \quad \mbox{and} \quad S_T = \left(\frac{3q\nu^5}{g^5k^4}\right)^{1/4}. \label{constflux}
\eqno{(\theequation{\mathit{a},\mathit{b},\mathit{c}})} 
$$
This system admits a steady state in which drainage exactly balances the material input. In the long-time limit, the depth of drainage greatly exceeds the height of the current $L \gg H$ and the current has the steady profile
\begin{equation}
H = \left[ R^2-R_N^2-2R_N^2\ln{(R/R_N)}\right]^{1/4}, \label{ssshape}
\end{equation}
plotted in figure \ref{ss}. The logarithmic singularity at $R = 0$ accounts for the finite flux there \citep{huppert-1982}. Balancing the external flux with the drainage flux, we find the steady-state extent
\begin{equation}
R_N(t\rightarrow\infty) = \pi^{-1/2} \approx 0.564. \label{sslength}
\end{equation}
We note that \citet{pritchard-2001} determined an analytical solution for the steady state of currents in a similar system. They considered a two--dimensional current fed by a constant source at the origin spreading through a porous medium of high permeability and underlain by a very thin low permeability porous layer. Comparing their figure 3 and equation (2.20) with our figure \ref{ss} and equation (\ref{ssshape}) highlights the importance of the specific system to the current behaviour. Whereas in our system, the steady-state profile of the current has an inflection point where the curvature switches from being positive near the source to negative near the nose, in their system the steady--state current surface is concave upwards near the nose.

\section{Numerical solution}

The full time evolution of the current height $H\left(R,T\right)$ and depth $L\left(R,T\right)$ is found numerically by integrating equations (\ref{NDH}) and (\ref{NDL}) on a uniform grid with spacing $0.001$. We compute the new drainage depth $L_{n+1} = L \left(R, T_{n+1}\right)$ from equation (\ref{NDL}) using the height $H_{n} = H\left(R,T_n\right)$ and depth $L_{n} = L\left(R,T_n\right)$ from the previous time step. We then use $L_{n+1}$ to compute the drainage velocity on the right-hand side of equation (\ref{NDH}) and solve for the height from equation (\ref{NDH}) using the control volume (or flux conservative) method in space and a Crank--Nicholson (semi-implicit) scheme in time \citep{patankar-1980}. In this computation, $\left(H_n\right)^3$ is our initial estimate for $H^3$ in the non-linear term on the left-hand side of equation (\ref{NDH}). We update the height as described above and use the new estimate for $H$ in the non-linear term, iterating this process until the updated value of $H$ converges. This converged value is $H_{n+1}$. Finally, we proceed to the next time step.

We introduce fluid into the current by assigning a constant flux at the left-hand boundary of the first control volume ($R = 0$) in the discretized equations. The right-hand boundary of our grid is impermeable to fluid flow and is positioned beyond the steady-state extent. The initial condition is an empty box with no fluid. We record the nose of the current as the position where $H(R)$ falls below a prescribed small tolerance: the height beyond this point is set to $0$. This condition is necessary because the drainage velocity is ill-defined when $L = 0$ and $H \not= 0$, as occurs near the current nose at the beginning of a time step.

We have tested the sensitivity of the numerical solution to the choice of grid spacing, time--step size and height tolerance, and found the results to be relatively insensitive to these parameters. Additionally, we have tested our numerical results for non-draining currents against those of \cite{huppert-1982}, and find good agreement for both the constant volume and constant flux cases.

\begin{figure}
\centering
\includegraphics[width = 4.5in]{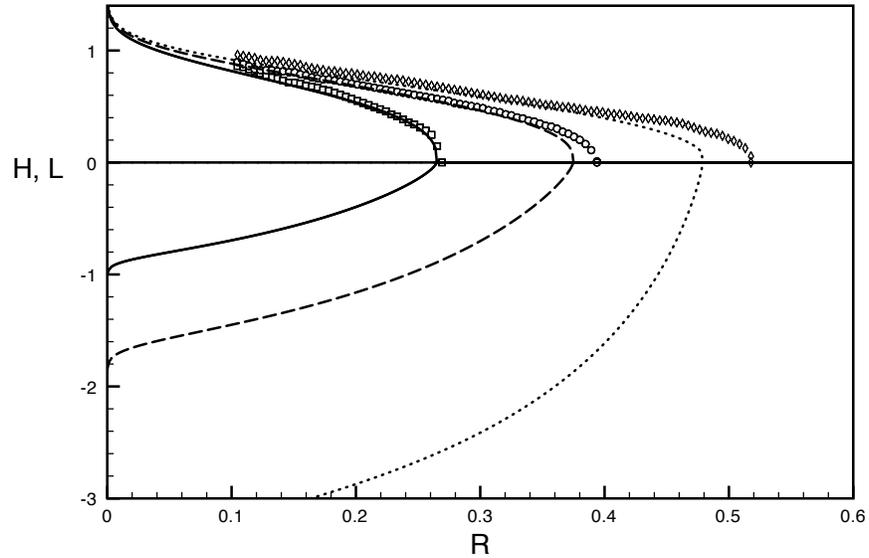}
\caption{Dimensionless height $H$ and depth $L$ profiles from the numerical solution with $\phi = 0.907$ (curves) and experiment 9 (symbols) at dimensionless times $T = 0.216$ (solid curve and $\medsquare$), $0.592$ (dashed curve and $\medcircle$), and $1.72$ (dotted curve and $\lozenge$). Data for $R \lesssim 0.1$ are not plotted as effects from the finite width and coiling instability of the fluid source are most pronounced there. See discussion of experimental errors in \S 4\label{numprof}}
\end{figure}

Figure~\ref{numprof} shows three calculated profiles (curves) at different times for a numerically simulated current with porosity $\phi = 0.907$, similar to that of our experimental setup. We note that, as the system approaches steady state, the volume of fluid residing within the porous medium becomes much larger than the volume in the current above the medium. The solid curve in figure~\ref{TheoryExpt} shows how the extent increases with time, approaching the steady-state value predicted by equation (\ref{sslength}).

\section{Experiments}

\begin{figure}
\centering
\includegraphics[width = 5in]{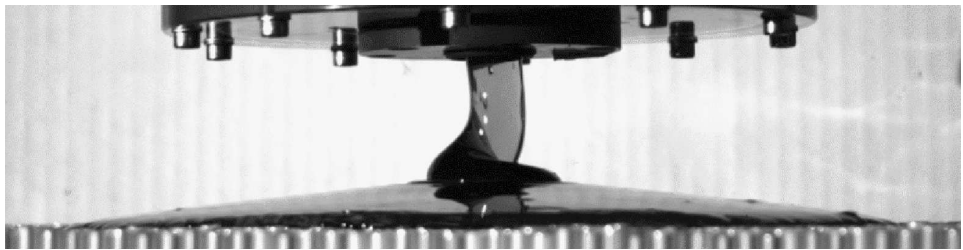}
\caption{Lyle's golden syrup spreading out over and draining into a bed of drinking straws. \label{ExptFig}}
\end{figure}

We performed a series of experiments using Lyle's golden syrup as the viscous intruding fluid and a bed of vertically oriented drinking straws of radius $r_s = 0.29 \pm 0.01$ cm as the underlying porous medium, as shown in figure~\ref{ExptFig}.  Lyle's golden syrup was used as the working fluid because its viscosity of $\nu \gtrsim 400\ \mbox{cm}^2 \mbox{s}^{-1}$ (as measured by a U-tube viscometer) results in currents with heights much greater than the surface topography of the porous medium. The simple geometry of the porous medium ensures strictly vertical drainage flow and allows for a comparison between the experimentally measured and the theoretically predicted permeability.

We measured the permeability of the porous medium by conducting drainage experiments in which syrup with $\nu = 453\ \mbox{cm}^2 \mbox{s}^{-1}$ and $\rho = 1.5\ \mbox{g}\ \mbox{cm}^{-3}$ was maintained at a constant height $h = 10\ \mbox{cm}$ above the porous medium within a large cylinder of radius $r_C = 5.75\ \mbox{cm}$ as it drained through the straws of length $l = 20\ \mbox{cm}$. By measuring the mass flux $dM/dt$ through the straws with a digital scale connected to a computer, we obtained the drainage velocity
\begin{equation}
w = \frac{g k}{\nu}\left(1+\frac{h}{l}\right) = \frac{dM/dt}{\rho \pi r_C^2},
\end{equation}
and consequently a measure of the permeability of the porous medium
\begin{equation}
k_{exp} = 6.36 \pm 0.04 \times10^{-3}\ \mbox{cm}^2.
\end{equation}
The uncertainty in this value comes from estimating $dM / dt$ from the measured mass versus time, which has an uncertainty of $\pm 0.02\ \mbox{g}\ \mbox{s}^{-1}$. This experimental value can be compared with the theoretical permeability for aligned capillary tubes given by \cite{bear-1972}
\begin{equation}
k = \frac{\phi r_s^2}{8} = 9.5 \pm 0.3 \times10^{-3}\ \mbox{cm}^2,
\end{equation}
where $\phi = \pi \sqrt{3} / 6 \simeq 0.907$ is the packing fraction of the straws for hexagonal close packing. We attribute the approximately $30 \%$ discrepancy between the measured and theoretical values to a slow leakage of golden syrup through the interstices between the straws and imperfections in the straw packing that produced a porosity not equal to that of a close packing. In the following analysis of the experimental data, we use the measured permeability.

For each experiment, a fixed flux of syrup was supplied at the origin from a reservoir maintained at a constant gravitational head. The mass flux was measured with a digital balance connected to a computer prior to the initiation of each experiment. The flux, viscosity and resultant scaling laws are summarised in table~\ref{ExptValues2} for each experiment.

\begin{table}
\begin{center}
\begin{tabular}{@{}ccccccc@{}}
Experiment & $q\ (\mbox{cm}^3 \mbox{s}^{-1})$ & $\nu\ (\mbox{cm}^2 \mbox{s}^{-1})$ & $S_H\ (\mbox{cm})$ & $S_V\ (\mbox{cm})$ & $S_T\ (\mbox{s})$ \\
1 & $1.06\pm0.01$ & $453$ & $8.78$ & $1.10$ & $80.0$ \\
2 & $4.18\pm0.03$ & $453$ & $17.4$ & $1.55$ & $113$ \\
3 & $2.09\pm0.01$ & $453$ & $12.3$ & $1.30$ & $94.7$ \\
4 & $9.83\pm0.01$ & $453$ & $26.7$ & $1.92$ & $139$ \\
5 & $6.74\pm0.01$ & $453$ & $22.1$ & $1.75$ & $127$ \\
6 & $2.31\pm0.01$ & $401$ & $12.2$ & $1.30$ & $83.4$ \\
7 & $1.27\pm0.01$ & $401$ & $9.05$ & $1.12$ & $71.9$ \\
8 & $2.00\pm0.01$ & $401$ & $11.3$ & $1.25$ & $80.4$ \\
9 & $6.11\pm0.01$ & $401$ & $19.8$ & $1.65$ & $106$ \\
10 & $7.37\pm0.01$ & $401$ & $21.8$ & $1.73$ & $111$ \\
11 & $6.63\pm0.01$ & $401$ & $20.7$ & $1.69$ & $109$ \\
12 & $20.35\pm0.03$ & $401$ & $36.2$ & $2.24$ & $144$ \\
\end{tabular}
\end{center}
\caption{Summary of the experimental parameters. For each experiment, the permeability was assumed to be $k = 6.36 \pm 0.04 \times10^{-3}$ cm$^2$. The uncertainty in viscosity is described in the text. \label{ExptValues2}}
\end{table}

\begin{figure}
\centering
\includegraphics[width = 4.5in]{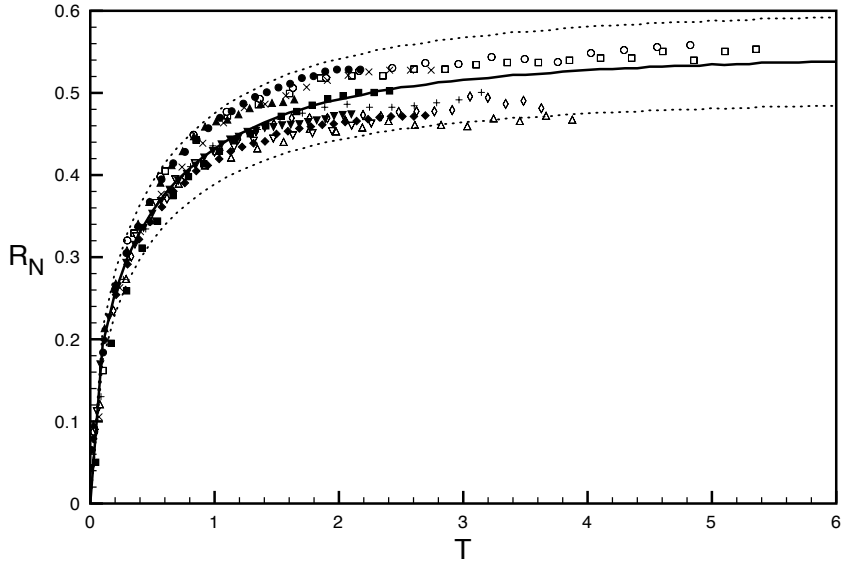}
\caption{The dimensionless radial extent $R_N$ versus dimensionless time $T$ for the experiments listed in table \ref{ExptValues2} and our numerical solution (solid curve). The dotted curves show the $\pm 10 \%$ error bounds. The symbols correspond to experiments: 1 ($\medsquare$), 2 ($\medcircle$), 3 ($\bigtriangleup$), 4 ($\diamond$), 5 ($\bigtriangledown$), 6 (+), 7 (x), 8 ($\filledmedsquare$), 9 ($\bullet$), 10 ($\blacktriangle$), 11 ($\filleddiamond$), and 12 ($\blacktriangledown$). \label{TheoryExpt}}
\end{figure}

Digital images of the side profile of each experiment were made at regular intervals (see figure~\ref{ExptFig}), and later analysed to obtain the radial extent and height profiles of each current. A comparison between the scaled radial extent of each current and the numerical solution to equations~(\ref{NDH})--(\ref{NDL}) is shown in figure~\ref{TheoryExpt}. The dotted curves represent a $\pm 10 \%$ error bound in $S_H$ applied to the numerical extent (solid curve). The error in $S_T$ is not represented in the plot. Uncertainty in $\nu$ and $k$ are the main contributors to the overall uncertainty, as the error in $q$ is less than $1\%$. Although the viscosity was measured regularly throughout the set of experiments, the large range of values obtained and the known large temperature dependence of the viscosity ($20\%$ per $^\circ$C) result in a large uncertainty in the actual viscosity. Due to this uncertainty in the viscosity and the discrepancy between the theoretical and measured permeabilities, we estimate the total error to be about $\pm 10\%$. Within the error bounds, the collapse of the scaled data and agreement with the numerical solution is good. We also obtained height profiles from the images of experiment 9, which are compared to the numerical profiles at the same scaled times in figure \ref{numprof}. For clarity error bounds are not plotted, but the uncertainty is again $\pm 10\%$. Although the finite width and coiling instability of the fluid source cause some discrepancy, the overall agreement is good.

\section{Discussion}

\begin{figure}
\centering
\includegraphics[width = 13cm]{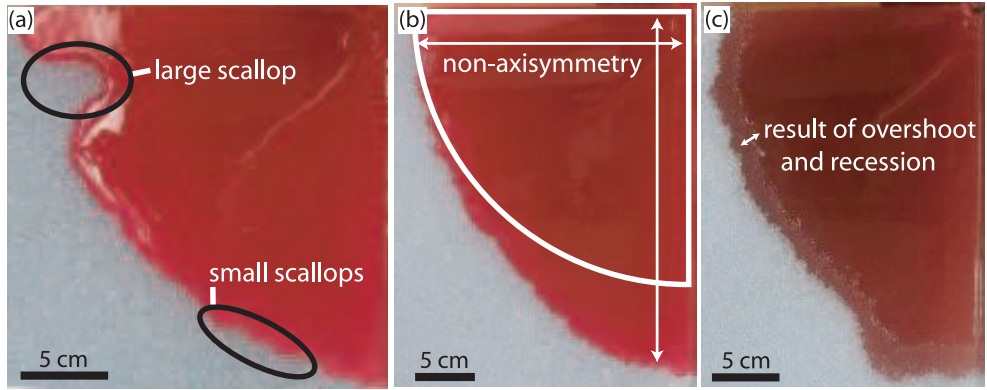}
\caption{Examples of non--ideal behaviour from three gravity currents using glycerin (dark) input at the upper right corner and spreading across beads (light); (a) large and small scallops at the leading edge; (b) non--axisymmetric spreading (the quarter circle is equiaxial); and (c) a region of beads over which the current advanced and subsequently receded. \label{badcurrent}}
\end{figure}

We have shown that for a fixed flux of golden syrup flowing across a bed of vertically aligned straws a simple model based upon lubrication theory and the drainage law of \cite{acton-2001} can describe the current propagation and steady state. In contrast, experiments conducted using glycerin as the working fluid and $\sim 3$ mm diameter spherical glass beads as the porous medium (detailed results not included here) exhibited non--ideal behaviour that violated a number of assumptions in our model. In particular, most currents had a scalloped front as the current propagated across the beads (figure \ref{badcurrent}\emph{a}), complicating measurement of the current radius. Many were also non--axisymmetric as shown in figure \ref{badcurrent}\emph{b}. Finally, all of the glycerin currents exhibited a maximum extent from which the current nose then retreated (figure \ref{badcurrent}\emph{c}).

We attribute these non--ideal behaviours primarily to a sensitive dependence on the characteristics and geometry of the underlying porous medium. For example, at the nose of the current the thickness is small and may be comparable to the surface topography of the porous medium. This could cause the front to stick on surface asperities, producing the scalloped edge. Additionally, the currents are sensitive to inhomogeneities in the bead packing, and thus the permeability of the medium, due to the strong influence of permeability on the drainage velocity in equation (\ref{depthevo}) and on the scaling laws in equation (\ref{constflux}). This may have contributed to the scalloped front and the non--axisymmetric propagation. The non--axisymmetry also could have arisen from a bead surface that was not sufficiently level. Although care was taken to level the surface, we cannot exclude this possibility. These hypotheses are supported by our experiments using golden syrup and straws, a level and uniform porous medium, and could be tested by conducting more experiments using, for example, glycerin and straws or smaller beads.

For the roll-back phenomenon, we have no simple explanation. However, we can rule out some possibilities. First, we verified that horizontal flow within and immediately above the surface of the porous medium was negligible as assumed in our model (no-slip condition). Powdered dye placed in several small piles on the bead surface along the path of the current was picked up by the draining fluid and carried purely vertically into the beads. Secondly, the geometry of the glycerin experiments afforded us a cross-sectional view of the current and draining fluid from which we observed a uniform saturation of the beads. This supports our assumption of a constant permeability, though as we could not observe the interior of the porous medium, we cannot completely rule out these effects in the bulk of the flow. Because the roll-back phenomenon was not observed in the golden syrup and straws system, we think that it is related to the specific combination of fluid and porous medium properties. Again, this could be tested with experiments involving different fluids and porous media. Finally, we note that no experimental evidence for a Rayleigh--Taylor instability was found at the lower interface of the current on the time scales over which the experiments were conducted. This observation implies that, at least here, vertical drainage is the dominant factor controlling radial spreading of the current.

The contrast between our experiments using glycerin and beads and those of \cite{acton-2001} using glycerin and beads in a linear geometry with a fixed fluid volume suggests that some characteristic of either the axisymmetric geometry or the fixed fluid input results in currents that are much more sensitive to the properties of the porous medium. For example, axisymmetric currents have a much longer front and therefore a larger nose area than linear currents. Therefore their spreading is more strongly influenced by surface roughness and the failure of our model assumptions near the nose. To explore these ideas further, we suggest conducting fixed flux experiments in the linear geometry and fixed volume experiments in the axisymmetric geometry using different porous media.

The sensitive dependence of propagation and drainage of the current on the spatial structure of the permeability and the surface topography suggests that further studies are needed to characterise fluid flow in these situations. Nonetheless, our model provides a simple framework to estimate the evolution of the current over time and the maximum extent at steady state for currents flowing over simple porous media.\newline

We wish to thank Mark Hallworth and Michael Patterson for their assistance with the experiments. This research was partially supported by the U.S. Department of Energy (DE-FG02-05ER15741). MJS acknowledges support from a U.S. National Science Foundation Graduate Research Fellowship. JSW gratefully acknowledges support from the Wenner-Gren Foundation, the Royal Institute of Technology, and NORDITA in Stockholm.

\bibliographystyle{jfm}

\end{document}